\documentclass[apj]{emulateapj}

\newcommand{\myemail}{schenker@astro.caltech.edu}
\newcommand{\lt}{\ifmmode\,<\,\else \,$<$\,\fi}
\newcommand{\kms}{\ifmmode\,{\rm km}\,{\rm s}^{-1}\else km$\,$s$^{-1}$\fi}

\newcommand{\magarc}{\ifmmode {{{{\rm mag}~{\rm arcsec}}^{-2}}}
             \else {{{mag}$~${arcsec}$^{-2}$}}
             \fi}

\newcommand{\yhst}{$Y_{105}$}
\newcommand{\jhst}{$J_{125}$}
\newcommand{\hhst}{$H_{160}$}

\newcommand{\xlya}{x$_{\mathrm{Ly}\alpha}$}
\newcommand{\lya}{$\mathrm{Ly}\alpha$}
\newcommand{\muv}{M$_{\mathrm{UV}}$}
\newcommand{\ewlya}{EW$_{\mathrm{Ly}\alpha}$}
\newcommand{\xhi}{x$_{\mathrm{HI}}$}
\newcommand{\fcgs}{erg cm$^{-2}$s$^{-1}$}

\shorttitle{A Reanalysis of the Lyman Alpha Fraction}

\begin{document}

\title{Line Emitting Galaxies Beyond a Redshift of 7: An Improved Method for Estimating
the Evolving Neutrality of the Intergalactic Medium}


\author {Matthew A Schenker\altaffilmark{1},
Richard S Ellis\altaffilmark{1}
Nick P Konidaris\altaffilmark{1},
Daniel P Stark\altaffilmark{2,3},
}

\altaffiltext{1}{Department of Astrophysics, California Institute of 
                Technology, MC 249-17, Pasadena, CA 91125; 
                \myemail}
\altaffiltext{2}{Department of Astronomy and Steward Observatory, University of Arizona, Tucson AZ 85721}
\altaffiltext{3}{Hubble Fellow}

\begin{abstract}

The redshift-dependent fraction of color-selected galaxies revealing Lyman alpha emission, \xlya\ has
become the most valuable constraint on the evolving neutrality of the early intergalactic medium.
However, in addition to resonant scattering by neutral gas, the visibility of Lyman alpha is also
dependent on the intrinsic properties of the host galaxy, including its stellar population, dust content and the
nature of outflowing gas. Taking advantage of significant progress we have made in determining
the line emitting properties of $z \simeq 4-6$ galaxies, we propose an improved method, based on
using the measured slopes of the rest-frame ultraviolet continua of galaxies, to
interpret the growing body of near-infrared spectra of $z>7$ galaxies in order to take into account these 
host galaxy dependencies. In a first application of our new method, we demonstrate its potential via a
new spectroscopic survey of $7<z<8$ galaxies undertaken with the Keck MOSFIRE spectrograph. Together
with earlier published data our data provides improved estimates of the evolving visibility of Lyman alpha,
particularly at redshift $z\simeq 8$. As a byproduct, we also present a new line emitting galaxy at a redshift $z=7.62$
which supersedes an earlier redshift record.  We discuss the improving constraints on the evolving neutral
fraction over $6<z<8$ and the implications for cosmic reionization.

\end{abstract}

\keywords{galaxies: evolution}

\section{Introduction}\label{sec:intro}

The transition from a neutral intergalactic medium (IGM) to one that is ionized, and therefore transparent to
ultraviolet photons, represents the latest frontier in our overall understanding of cosmic history. In addition to
determining when this `cosmic reionization' occurred, a key question is the role of star-forming galaxies in
governing the process. Structure in the polarization of the cosmic microwave background suggests
the reionization process occurred within the redshift interval $6<z<20$ \citep{Hinshaw2013a} and
deep infrared imaging with Hubble Space Telescope has provided the first opportunity to conduct a census of galaxies
during the latter half of this period \citep{Ellis2013a, Oesch2013a}. Recent progress in this area has
been reviewed by \cite{Robertson2013a} and \citet{Bromm2013a}.

In the absence of significant numbers of high redshift QSOs or gamma ray bursts, the most immediately
available probe of the evolving neutrality of the IGM beyond $z \simeq 6-7$ is the visibility of the Lyman alpha (\lya) 
emission line in controlled samples of color-selected galaxies. Although a prominent line in
star-forming galaxies at $z \leq 6$, as Ly$\alpha$ is a resonant transition, it is readily suppressed by neutral
gas, both in the host galaxy and, if present, in the surrounding IGM.  First proposed as a practical experiment 
using Lyman break galaxies (LBGs) by \citet{Stark2010a}, the idea followed earlier theoretical work by \citet{Miralda-Escude2000a}, 
\citet{Santos2004a} and others.

Ground-based near-infrared spectroscopic surveys have now targeted various samples of color-selected
Lyman break galaxies over $6<z<8$ allowing the construction of a redshift-dependent
\lya\ fraction, \xlya , which falls sharply from a value of $\simeq$ 50\% at $z\simeq 6$
\citep{Stark2010a} to less than 10\% at and beyond $z\simeq 7$ \citep{Pentericci2011a,Schenker2012a,
Ono2012a, Treu2012a,Pentericci2014a}. Although converting this downturn in the visibility of the line into the volume fraction of
neutral hydrogen, \xhi, is uncertain \citep{Bolton2013a}, the prospects for improving the statistics of
this test are promising given the arrival of multi-object instruments such as MOSFIRE on the Keck 1 telescope
\citep{McLean2012a}.

So far, this important measure of late reionization has been applied by adopting an empirical description
of the demographics of Ly$\alpha$ emission in LBGs, parameterized according to the equivalent width (EW) distribution for
various UV luminosities over the redshift range $3<z<6$ when the Universe is fully ionized. The trend is
then extrapolated to higher redshifts in the form of a `no evolution' prediction with the aim of rejecting this
prediction at some level of significance (e.g. \citealt{Schenker2012a}). As we discuss here, this method, 
now widely used, has several disadvantages. Recognizing these and noting the spectroscopic and optical and 
near-infrared imaging data of LBGs over $3<z<6$ has improved in scope and quality, in this paper we 
adopt a more physically-based approach to the visibility of Ly$\alpha$ in the vicinity of the host galaxy. Our new
approach aims to predict its visibility in a high redshift galaxy on the basis of its measured UV continuum
slope that, in turn, contains information on the dust content, and stellar population, which both directly
influence the strength of any Ly$\alpha$ emission.  
This approach has the distinct advantage that, for the new $z>7$ samples being studied with MOSFIRE and 
other spectrographs, composite UV slopes for the population are usually available so that unnecessary extrapolation 
can be avoided.

The present paper is concerned with describing this improved Ly$\alpha$ fraction test and applying it 
to the first comprehensive set of spectroscopic data emerging from MOSFIRE. In addition to incorporating 
the earlier surveys conducted with Keck \citep{Schenker2012a,Ono2012a,Treu2012a}, and 
FORS2 on the VLT \citep{Pentericci2011a}, we present the first results from a survey of high quality Ultra Deep Field (UDF) targets 
which provides a valuable extension of the aforementioned studies.
As part of this survey, we demonstrate a new Ly$\alpha$-emitting galaxy at a redshift $z$=7.62 extending once
again the frontier of spectroscopically-confirmed HST sources.

A plan of the paper follows. In Section 2, we introduce our new method for the Lyman alpha fraction test.  
Section 3 introduces the new compilation of $3<z<6$ data drawn from our now completed Keck spectroscopic
survey \citep{Stark2014a}, and Section 4 contains an analysis of this data in the context of our new method. 
In Section 5 we introduce our new MOSFIRE data and apply our new method to both this data and that obtained
earlier. 

\section{Lyman Alpha Fraction Test - A New Approach}

Although the traditional Ly$\alpha$ fraction test as first proposed by \citet{Stark2010a} has already provided 
meaningful constraints on the evolution of the IGM beyond $z \sim 6.5$, there are two limitations in the current
methodology. Firstly, as inferring the presence of neutral gas in the intergalactic medium represents a 
differential measurement, it is necessary to assume a form of the distribution of the equivalent widths 
of Ly$\alpha$ emission \textit{unprocessed by the IGM} for the galaxies at $z \geq 7$. Comparing this to the 
observed distribution allows
the extinction imposed by the IGM, and through the application of theoretical models, the IGM neutrality
to be derived. The current methodology splits the sample into UV-luminous and UV-faint bins, and tracks
the Ly$\alpha$ fraction in each bin as a function of redshift.  The IGM unprocessed distribution at $z \sim 7$ is then
assumed to either be that observed at $z \sim 6$ for each bin, or a linear extrapolation of the $3 < z < 6$
data. However as we cannot observe the sample at $z \geq 7 $ in the absence of a neutral IGM, we can never know
which, if either, of these assumptions is correct.

Secondly, the EW distribution used to predict the observable \lya\ distribution has been characterized in many
different ways, including an exponential \citep{Dijkstra2011a}, a direct histogram \citep{Schenker2012a}, a half gaussian \citep{Treu2012a} and 
a half gaussian with a constant probability tail \citep{Pentericci2011a}. Though the distributions are largely similar,
no detailed comparison has been performed to determine which one optimally represents the $3 < z < 6$ data. We 
perform this in the context of assembling our model in Section \ref{sec:analysis}.

The most fundamental question, however, is whether the {\it rest-frame UV luminosity} is the optimum parameter to predict
the visibility of Ly$\alpha$ in the absence of any suppression by a partly neutral IGM. The approach, based
on correlations first noted by \citet{Shapley2003a}, was adopted by \citet{Stark2010a} as  \muv\ can be readily 
determined from the available photometry of distant galaxies together with a photometric redshift. However, 
\muv\ is likely to be a coarse predictor of the Ly$\alpha$ EW as it ignores second order parameters such
as metallicity, the stellar initial mass function and dust content. 
 
The {\it UV continuum slope} is a more natural choice as a basic variable as it encodes each of these physical
properties \citep{Meurer1999a}.  Lower metallicity and hotter stars produce more ionizing photons per unit UV 
continuum, thus driving \ewlya\ upwards. Dust very efficiently absorbs \lya\ photons given their large effective 
path lengths from the many scatterings required to escape an HII region. These changes also result in a bluer or
redder UV continuum slope, respectively. Thus, as the UV slope reflects
more of the parameters that likely govern \ewlya\, compared to \muv, we should expect it to be a more robust predictor 
of the visibility of the line in high redshift samples.

Until recently, determining the UV continuum slope was only possible for a restricted subset of $z<4$ 
$B$-dropouts. \citet{Stark2010a} showed that within this subset, strong \lya\ emitting galaxies have bluer UV continuum 
slopes than their non-emitting counterparts, but as there existed no high-quality infrared photometry in the 
GOODS fields at this time, it was necessary to parametrize distributions at higher redshift by their absolute magnitude.
However, in addition to the now completed Keck spectroscopic survey of LBGs over $3<z<6$ \citep{Stark2010a,Stark2014a}, the 
CANDELS HST imaging program (GO 12444-5, PI: Ferguson/Riess/Faber) provides the necessary data to explore the potential
of the UV continuum slope as a predictor for \ewlya. The addition of \yhst, \jhst, and \hhst\ photometric data enables the 
derivation of accurate UV continuum slopes for catalogued galaxies, given for each source there is a minimum of 3 
broad-band filters longward of the Lyman break.  As such UV continuum slopes are now available for the growing 
body of $z>7$ photometric galaxy samples(e.g. \citealt{Dunlop2013a, Bouwens2013a}), we can realize a 
Ly$\alpha$ fraction test that overcomes several of the issues associated with the earlier approach.

In the following, we discuss the new data for the Keck $3<z<6$ spectroscopic sample \citep{Stark2014a} and analyze 
it in the context of a distribution function based upon the observed UV continuum slopes of the population.
We then apply the method to an updated sample of near-infrared spectroscopic data beyond $z\simeq6$.

\section{Improved Post-Reionization Data}

\subsection{DEIMOS/FORS2 Spectroscopy}

As discussed in \citet{Stark2010a,Stark2011a}, the $3<z<6$ LBG candidates which form the basis of the
post-reionization sample were targeted in the GOODS-N and S fields using the DEIMOS spectrograph on
the Keck 2 telescope. The final catalog is being prepared for release by \citet{Stark2014a}. By retrospectively
applying the same photometric selection criteria, a spectroscopic sample in the GOODS-S field using the FORS2 
spectrograph on the ESO Very Large Telescope was added (\citet{Vanzella2009a} and references therein). 
Full details of these spectroscopic campaigns can be found in the above referenced articles.

The GOODS-N sample consists of 393 LBG candidates targeted with DEIMOS observed over the course
of 2008-2010.  The targets include $B$-,$v$-, and $i$-drops, and the spectroscopically-confirmed sample spans
a redshift range of $3.33 < z < 5.99$.  Typical 10$\sigma$ limiting Ly$\alpha$ fluxes for these targets ranged between
1.0-1.5 $\times 10^{-17}$ \fcgs.

The complementary FORS2 campaign \citep{Vanzella2009a} targeted 214 LBG candidates in GOODS-S between
2002 and 2006.  These targets were, on average, brighter than those studied at Keck (see \citealt{Stark2010a}, Figure 2),
and the confirmed galaxies span a redshift range $3.19 < z < 6.28$.  In total, the sample comprises 607 galaxies, 269 
of which are spectroscopically confirmed.  We direct the reader to \cite{Stark2014a} for further details.

\subsection{Photometry}

The primary advance in our analysis of the equivalent width distribution of Ly$\alpha$ in the above
spectroscopic sample relates to the combination of the earlier HST ACS optical imaging data with new, deep WFC3/IR
near-infrared data critical to assessing how \lya\ emission correlates with the measured UV continuum slope.  To
reliably bring together the various imaging datasets, it is necessary to account for the significantly different 
point-spread functions (PSFs) between the ACS (FWHM $\sim$ 0.09'') and WFC3/IR (FWHM $\sim$ 0.16'') instruments. 
In the GOODS-S field, we utilized the published, PSF-matched catalog of \cite{Guo2013a} which uses the publicly 
released v2.0 ACS and v1.0 WFC3 images, constructs stellar profiles to derive the PSFs in order to convolve the 
higher-resolution, shorter wavelength data, and performs isophotal photometry on the smoothed images.

For GOODS-N, we also utilized the publicly released v2.0 ACS images, but as no CANDELS WFC3/IR mosaic
was released at the time of this analysis, we constructed our own.  The first 13 (out of 18) epochs of
GOODS-N observations, released as individual, v0.5 mosaics, were combined using the routine SWARP \citep{Bertin2002a}
with individual weights assigned according to the exposure time.  PSF matching was implemented using the \textit{ColorPro} 
program \citep{Coe2006a}. A PSF was constructed for each filter by shifting and stacking $\sim 20$ bright unsaturated stars.  
All objects were detected using the \hhst\ image, and colors were determined using matched isophotal apertures 
after degrading the resolutions of all other images to that of the \hhst\ image.

\section{Analysis}
\label{sec:analysis}

\subsection{Lyman alpha and the UV continuum}

The basis of our analysis relies on accurate determinations of both the Ly$\alpha$ equivalent widths (\ewlya), and 
ultraviolet slopes of our sample.  Thus, we detail here the methodology used in determining both 
these quantities for use in our analysis.

In order to measure the UV slope for each object in our sample, we first used a custom photometric redshift code to 
determine the approximate redshifts of those galaxies without spectroscopic confirmation. The code fits a suite of synthetic fluxes 
from the Bruzual-Charlot (BC03, \citealt{Bruzual2003a}) spectra to the available photometry.  To determine the 
best-fit redshift, we marginalized across all other parameters (mass, dust extinction, and age), and used the maximum 
likelihood value. 

We measure the UV continuum slope using the $\beta$ formalism first introduced by \citet{Calzetti1994a}, where
the flux is parametrized as $f_{\lambda} \propto \lambda^{\beta}$. In our fitting process, we include
photometric filters with central rest wavelengths within the range defined by \citet{Calzetti1994a}, 
$1350 < \lambda /$\AA $< 2600$. This range is also similar to that used 
previously in the literature \citep{Bouwens2013a,Rogers2013a}. As in \citet{Bouwens2013a}, we use the effective filter 
wavelengths appropriate for a $\beta = -2$ spectrum, since the measured UV slopes in our sample generally populate
the $-2.5 < \beta < -1.5$ range, and an error floor of 0.05 mag, or $5\%$, for all filters.

A grid of power law slopes with $-3.5 < \beta < 0.5$ and $\Delta \beta = 0.01$ was fit to the observed photometry, and the
relative likelihood of each computed using $p(\beta_i) \propto \mathrm{exp}(-\chi^2 / 2)$, appropriate for gaussian
distributed errors.  This allows us to construct a likelihood curve for the UV slope of each galaxy, central
to the fitting method we describe later.  After the fitting,
the image cutouts, photometry, and resulting $p(\beta)$ were manually inspected for each galaxy, flagging and removing
objects with clearly deviant photometry or incorrect solutions. After this process, 297 of our sample of 393 galaxies in 
GOODS-N, and 154 of our 214 galaxies in GOODS-S remained. Galaxies removed were typically those adjacent to bright
objects, for which accurate photometry could not be assured, and faint, distant $z \geq 5$ galaxies that appeared in
the shallower CANDELS-Wide field for which accurate UV slopes could not be determined.

Measurements of the \lya\ EW were taken from \cite{Stark2014a} with errors computed using both the $1\sigma$ flux errors 
in the spectrum, as well as errors in \muv, added in quadrature to produce a likelihood curve for $p($\ewlya$)$.  
For cases where \lya\ was not spectroscopically detected, we assume one of three cases: (1) the line flux falls 
below the $10\sigma$ limit (typically 1.0-1.5 $\times 1.0^{-17}$ \fcgs), (2) the line emission, though brighter than the 
10$ \sigma $ limit is missed, due to poor sky subtraction or obscuration by skylines, or (3) the object is a contaminant
outside the expected redshift range.  We discuss the implementation of this approach in Section 4.3.1.

\begin{figure*}
\begin{center}

\includegraphics[width=0.6\textwidth]{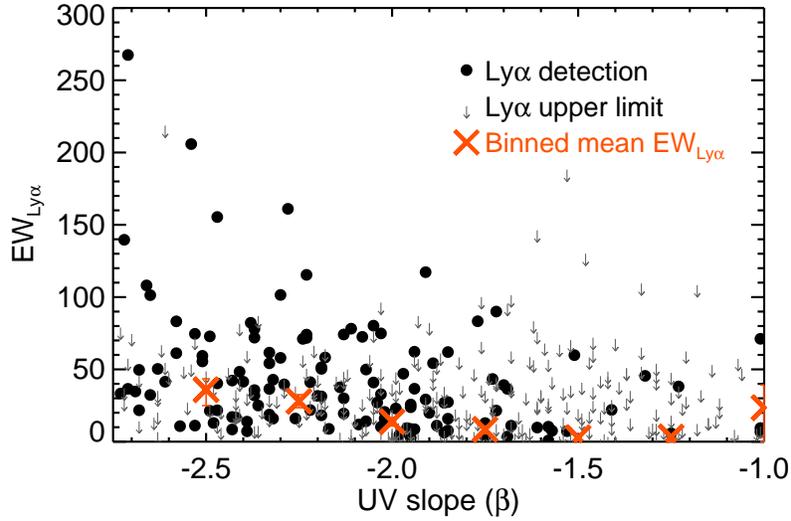}

\caption{\label{fig:lya_vs_beta}  Compilation of our entire GOODS catalog of Ly$\alpha$ equivalent 
widths as a function of UV slope, $\beta$.  Red triangles show the average EW, binned in steps of 
0.25 in $\beta$, displaying a strong increase toward bluer slopes.  This dataset forms the basis of
our predictive model for Ly$\alpha$ emission incidence as a function of UV slope.}
\end{center}
\end{figure*}

With these results in hand, we now provide the basic evidence that the UV continuum slope of a galaxy, $\beta$, is a reliable
predictor of its Ly$\alpha$ EW.  In Figure \ref{fig:lya_vs_beta}, we plot the best fit $\beta$ for each galaxy in our
final sample against either its \ewlya, or its 10$\sigma$ upper-limit, if a line is not detected. The red crosses
denote the mean \ewlya in bins spanning $\Delta \beta = 0.25$, where undetected objects are set to have a value of 0.
Error bars for each bin are calculated by bootstrap sampling. 

As shown from earlier work by \citet{Shapley2003a} and \citet{Stark2010a}, a clear trend of increasing mean \ewlya\ with 
bluer (more negative) UV slopes is visible.  With our large spectroscopic sample, this can be seen directly through
the measures of individual objects, rather than via stacked spectra or consideration of average $\beta$ values. 
Encouraged by this trend, we now develop a model that can predict the probability distribution of \ewlya\ 
given a measured value of $\beta$, for example for a $z>7$ galaxy. At the end of this section, we will also use this model
to show that the UV slope is a more reliable predictor of \ewlya\ than the absolute UV magnitude.

\subsection{The UV slope-dependent EW distribution}

We now seek to establish a formalism to predict the probability distribution for \ewlya\ given a particular measured UV slope.  
As our goal is to improve the accuracy of the Lyman alpha fraction test, we first consider those galaxies with blue UV slopes,
likely to have strong Ly$\alpha$ emission before processing through the IGM. 

\subsubsection{Equivalent Width Distributions for a Fixed UV Continuum Slope}

As an illustration of our new method we begin by examining the distribution of the Ly$\alpha$ EW for a fixed
UV continuum slope, $\beta$. A natural choice is $\beta = -2.3$ given that \citet{Bouwens2013a} find that 
faint (\muv $\geq -19$), high-redshift ($z \geq 6$) galaxies have slopes that asymptote to this value. Near-IR 
spectrographs such as MOSFIRE have begun to target these galaxies in earnest (this work, \citealt{Finkelstein2013a}, 
\citealt{Treu2013a}), and it is becoming increasingly important to characterize their expected IGM unprocessed \lya\ emission.  
To construct a sample of galaxies for this task, we limit our overall sample of 468 galaxies with $3.19 < z < 6.28$ to 
those galaxies with a best fit value within $\Delta\beta = 0.25$ of $-2.3$, resulting in a total of 131 objects.

We now require a model to represent the IGM unprocessed EW distribution.  In Appendix A, we review four such options
 using the methodology outlined here, and find that a lognormal distribution provides a significantly better fit 
 than any of the others.  In this case the natural logarithm of \ewlya\, obeys a normal distribution.  The two relevant 
 parameters of the distribution, $\mu$ and $\sigma$, are typically referred to as the {\it location parameter} and 
 {\it scale parameter}, respectively. They denote the mean of the natural log of \ewlya\  and its variance.  However, 
 while the median of the distribution is given, as might be expected, by exp($\mu$), the mean is slightly larger at exp($\mu + \sigma^2/2$). 
A third parameter, $A_{em}$, determines the fraction of galaxies that have \ewlya\ $> 0$, as there is no reason a priori
to expect all galaxies to display \lya\ in emission. The resulting distribution can be written as:

\begin{eqnarray}
p(EW) &=& A_{em} \times \frac{1}{\sqrt{2\pi}\sigma \mathrm{EW}} \mathrm{exp}(-\frac{(\mathrm{ln(EW)-\mu})^2}{2 \sigma^2}) \nonumber \\
&+& (1.0-A_{em}) \times \delta(EW)
\end{eqnarray}

\begin{figure*}
\begin{center}

\includegraphics[width=0.8\textwidth]{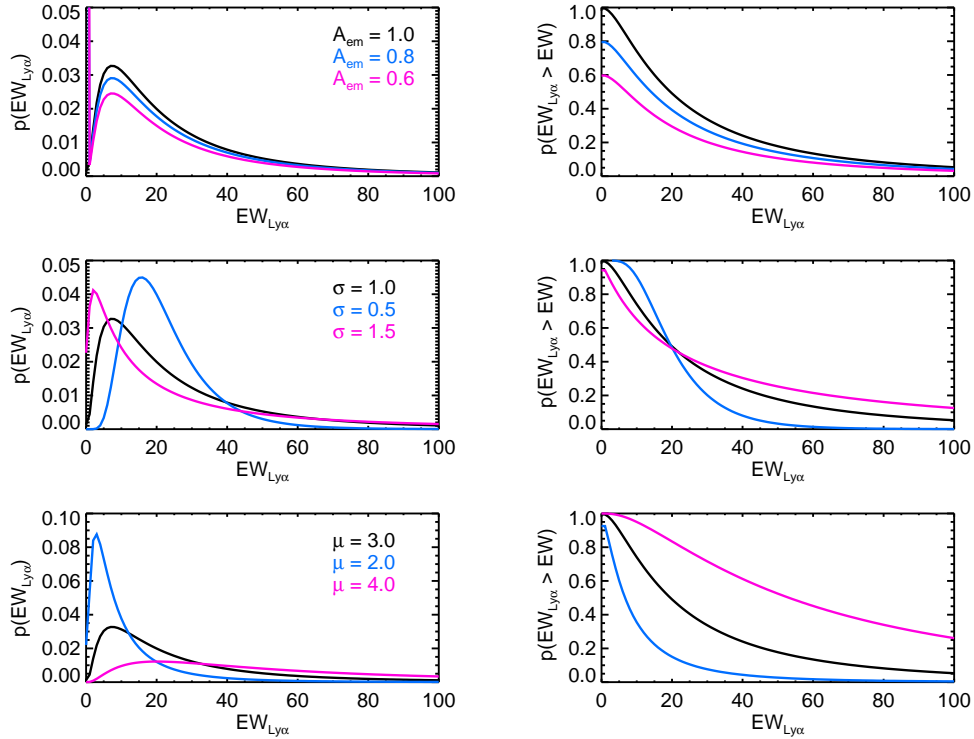}

\caption{\label{fig:longnorm_example} Example curves for how our lognormal model of \ewlya\ distribution
varies with each parameter. Left: Probability distributions for \ewlya. In each panel, the black curve has the same
parameter values: $\mu = 3.0$, $\sigma = 1.0$, $A_{em} = 1.0$. From top to bottom, the two colored curves 
each display the affect of a change in a single parameter on the distribution. Right: Complementary cumulative
distribution functions for the same parameters used in each left panel. This method of display
is especially useful, as the Lyman alpha fraction, \xlya, for any \ewlya\ can simply be read off the plot by
finding the value of the curve at the desired \ewlya\ along the x-axis.}
\end{center}
\end{figure*}

In Figure \ref{fig:longnorm_example}, we illustrate how the \ewlya\ probability distribution
function, $p$(EW), and the \lya\ fraction, \xlya, change as these parameters are varied.

We now describe the Bayesian formalism we developed to evaluate the likelihood of the underlying parameters for
our lognormal distribution, and determine which provides the best fit to the data. The entire set of spectroscopic \lya\ observations 
is denoted as \textbf{Obs}; this contains the information for observations of each individual galaxy, Obs$_i$. We can then 
denote the parameters for the model being fit as $\theta \equiv [\mu,\sigma,A_{em}]$.
Our overall goal is to determine the probability distributions for the underlying parameters of each model,
given our observations, i.e. $p(\theta | \textbf{Obs})$.  Using Bayes' theorem, we can rewrite this as

\begin{equation}
p(\theta | \textbf{Obs}) \propto p(\theta) \times p(\textbf{Obs} |\theta)
\end{equation}

Here, $p(\theta)$ represents our uniform priors for the underlying parameters, while the term on the right hand side represents the 
probability of our observations given the model parameters.  For any single object in which we measure a definite \ewlya, this
posterior probability can be expressed as:

\begin{equation}
p(Obs_i | \theta) = \int_0^\infty p(EW_{Obs,i}) p(EW | \theta) dEW
\end{equation}

In the case of an object for which \lya\ remains undetected above our ($10\sigma$) limit, we compute the posterior 
probability as:

\begin{eqnarray}
p(Obs_i |\theta) &=& p(EW < EW_{10\sigma} |  \theta) \nonumber \\
&+& p(EW > EW_{10\sigma} |  \theta) \times C_1 + C_2 
\end{eqnarray}

Here, the first term represents the probability that the object intrinsically posses an \ewlya\ below our detection
limits, while the $C_1$ term takes into account incompleteness in the sample (caused by skylines
or, in some cases, poor background subtraction).  Contamination by low redshift sources is taken into account through the final
term, $C_2$.  We assume modest values for our contamination terms of $C_1$ = $C_2$ = 0.05, motivated by 
the completeness simulations of \cite{Stark2010a}, and the lack of low-redshift interlopers found in other spectroscopic follow-up 
surveys \citep{Pentericci2011a}.  The full posterior distribution for the parameters can then be computed by multiplying the 
individual posterior probabilities for each object.  This allows us to infer the most likely values on parameters, as well as their 
marginalized and un-marginalized errors.  

\begin{figure*}
\begin{center}

\includegraphics[width=0.5\textwidth]{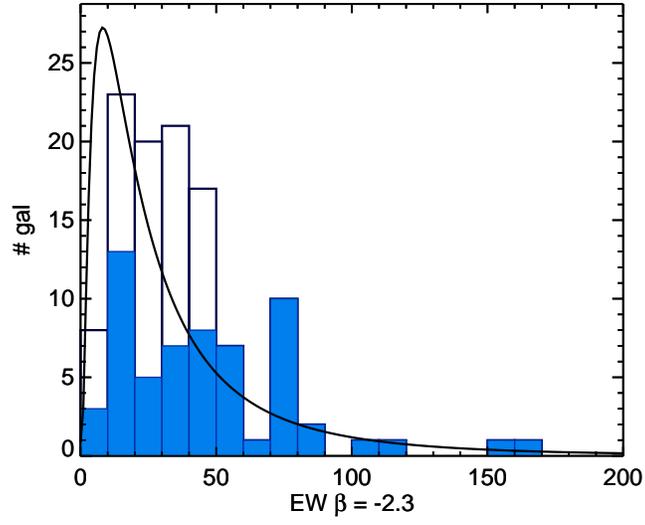}

\caption{\label{fig:b23_hist} Histogram of our observed \ewlya\ detections (solid blue) and upper limits
(unfilled black) for galaxies with best fit $\beta = -2.3 \pm 0.25$. Overplotted in solid black is our best
fit lognormal model as described in Section 4.2.1.}
\end{center}
\end{figure*}

We display the best-fit distribution for our sample of 131 galaxies with $3.19<z<6.28$, overplotted on a histogram of
their \ewlya\ detections and upper limits in Figure \ref{fig:b23_hist}.  The best fit parameters  are $\mu = 2.7 \pm 0.7$,
$\sigma = 1.4^{+0.9}_{-0.5}$, and $A_{em} = 1.0^{+0.0}_{-0.4}$. As we show in Appendix A, this formalism
is the best fit to our post-reionization data.

\subsubsection{A Generalized Approach}

Now that we have determined the distribution which best fits the data at the key UV slope value of $\beta \sim -2.3$,
we proceed with a more appropriate goal of determining the EW distribution across all values of $\beta$.
Although faint galaxies at $z \geq 6$ may asymptote to  $\beta \sim -2.3$, the UV-bright galaxies almost certainly do not \
\citep{Bouwens2013a}.  In order to fully leverage the \xlya\ test for the more luminous objects, we must use a model that 
determines the EW distribution across a wide range of $\beta$.  

To achieve this goal, we extend the Bayesian formalism introduced above.  The differences are twofold.  Firstly,
we now include our entire sample of spectroscopically observed galaxies when fitting, rather than just those
narrowly clustered around a particular value of $\beta$. Secondly, we must reconsider the nature of Eqn. 1
since it is clear that the EW distribution varies as a function of $\beta$ from Figure \ref{fig:lya_vs_beta}.

It is most reasonable to consider that the {\it location parameter}, $\mu$, varies with $\beta$ since it is
this parameter that governs the redistribution of EWs (see Figure 2). For convenience we assume that $\mu$
varies linearly with $\beta$, whence $\mu(\beta) = \mu_a + \mu_s \times (\beta + 2)$, where $\mu_a$ represents the 
location parameter at $\beta = -2.0$. Prior to selecting this model, we performed similar fits to a narrow range in 
$\beta$ as in the above section, with different ranges in $\beta$, and found that while the best fit for $\mu$ varied strongly 
with the UV slope, both $\sigma$ and $A_{em}$ did not.  

\begin{figure*}
\begin{center}

\includegraphics[width=0.5\textwidth]{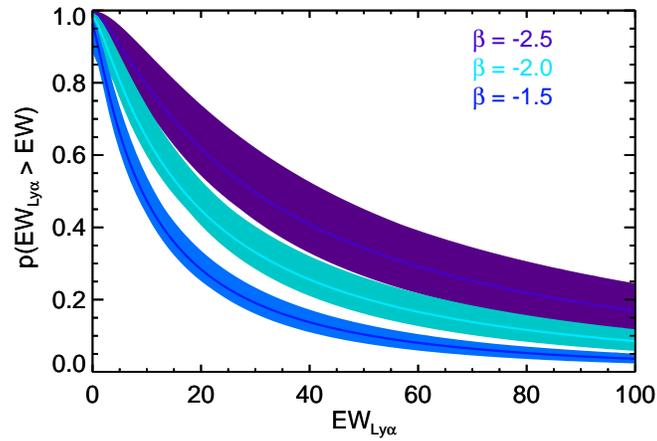}

\caption{\label{fig:pew_cdf} Inverse cumulative distribution functions of \ewlya\ for our best fit model, plotted as 
a function of $\beta$.  For a desired $\beta$, the Lyman alpha fraction, \xlya, for an arbitrary
equivalent width is defined by the y-value of the curve at the given \ewlya.}
\end{center}
\end{figure*}

Figure \ref{fig:pew_cdf} shows a representation of the product of the resulting formalism, while we
detail the full results in Appendix B. Because our model treats
the \textit{location parameter} as a linear function of $\beta$, we can generate an EW probability distribution
function for any UV slope, although our sample only provides meaningful constraints in the observed range $-2.5 < \beta < -1.25$.
In this figure, we plot 3 examples and their associated errors. Of course, a measured
UV slope for a high redshift galaxy has some error uncertainty, and thus its own probability distribution, $p(\beta)$. 
To obtain $p(EW)$ given our model, and an observation of $\beta$, we can marginalize over $\beta$ for each 
galaxy in the sample:

\begin{equation}
p(EW_i) = \int p(EW|\beta) p(\beta_i) d\beta
\end{equation}

\subsection{UV Slope versus UV Luminosity}

We now return to one of the assumptions motivating this paper.  Does parametrizing the likelihood of \lya\ emission
via the UV slope represent a statistically better option than the combination of absolute UV magnitude and redshift used in 
previous high-redshift \lya\ studies?  Since we can measure $\beta$, \muv, \ewlya\, and a photometric or spectroscopic
redshift for 451 objects in our spectroscopic sample, we can directly address this question. 

In the widely-used UV luminosity method, 
the dependence on \muv\ is handled in a discrete manner, assigning galaxies into one of two bins, depending on whether they are greater
or less than \muv\ $= -20.25$. \xlya\ is also calculated in discrete bins at $z = 4,5,$ and 6. For purposes of comparison, we use the
Bayesian formalism developed in this paper to generalize this to a continuous model. To do so, we alter the definition of $\mu$ from
$\mu(\beta) = \mu_a + \mu_s \times (\beta + 2.0)$ to $\mu(M_{\mathrm{uv}},z) = \mu_a + \mu_{s,M_{\mathrm{uv}}} \times (M_{\mathrm{uv}} + 19.5)
+ \mu_{s,z} \times (z - 4.0)$. This thus replaces the linear dependence of the location parameter, $\mu$, on UV slope, with a bivariate
linear dependence on UV magnitude and redshift. We then use the same fitting process to determine the optimal values for all
parameters in the model.

To compare how well each of these two models fits the available data, we use the Bayesian evidence ratio, or Bayes factor. The evidence is a 
measure of how likely the data are, given the model, and can be expressed as an integration of the likelihood function over all
possible values for each parameter in the model

\begin{equation}
E = \int p(\theta|\textbf{Obs}) d \theta
\end{equation}

Evaluating this for both models yields a significant gain via a ratio of $E_{\beta} / E_{M_{\mathrm{uv}},z} = 29$, showing convincing 
preference for the model based on $\beta$ compared to the earlier method.

\section{First Application to Data within the Reionization Era}
\label{sec:xlya_application}

Although the body of spectroscopic data targeting galaxies beyond $z \simeq 6$ in the reionization
era remains sparse, it is growing rapidly, particularly through the advent of multi-slit near-infrared
spectrographs such as MOSFIRE. Thus we are encouraged to apply our new
method to such data. In addition to collating earlier relevant data available in the literature,
we also present the first results from our new survey using MOSFIRE.

\subsection{A New MOSFIRE Survey}

As part of a long term survey targeting $z > 7$ galaxies using the MOSFIRE spectrograph on the Keck I 
telescope, we have secured deep spectroscopic observations in both November 2013 and
March 2014 targeting two different fields. One represents distant sources in a deep HST blank field with accurate
photometric redshifts and the other targeted gravitationally-lensed sources with extensive multi-band
photometry.

\subsubsection{The GOODS-South / Ultra Deep Field }

On the night of Nov 5, 2013, we secured a total of 3.5 hours exposure in the Hubble
Ultra Deep Field region of GOODS-South. Observations were taken using 0.7'' slits through intermittent high cirrus 
cloud and $\sim0.8$'' seeing. A total of 16 $z > 7$ candidates were included on this mask selected from an initial 
list of $z > 6.5$ targets from our UDF12 campaign and the GOODS-S field \citep{Schenker2013a,McLure2013a}, 
augmented with two additional $Y$-drops outside the UDF proper from \citet{Oesch2012a}.  We used the 
photometric redshift code, described in Section \ref{sec:analysis}, to compute a redshift probability distribution for each object.  

The UDF 2012 dataset\citep{Koekemoer2013a} (GO 12498, PI; Ellis) offers many distinct advantages for this first application of our method. 
Foremost, by virtue of the extraordinarily deep optical and F105W data available, contamination by foreground
objects, as determined by the photometric scatter simulations in \citet{Schenker2013a} is $\sim 3 \%$, down to the 
$J_{125} = 28.6$ limit of our targets. This contrasts with the $\geq 10 \%$ contamination rate affecting galaxies in the CANDELS fields \citep{Oesch2012a}.  
Secondly, as a result of a strategic deployment of near-infrared filters, our UDF 2012 candidates have better defined 
redshift probability distributions, allowing us to more confidently exclude the possibility of \lya\ emission in the event of a 
non-detection.  The median 68$\%$ confidence interval in photometric redshift for the UDF 2012 objects on our 
MOSFIRE mask is smaller by 
$\Delta z = 0.2$ compared to our GOODS-S targets (and most likely to earlier published blank field surveys - see \S5.3).

The final target selection for this aspect of our survey was arranged to formally maximize the expected number of detected lines, 
and thus our leverage in calculating \xlya.  As a first attempt,  we used the $z \sim 6$ histograms of \cite{Stark2011a} 
to predict the distribution of \ewlya for each object, as a function of its UV magnitude.  The fractional number of 
expected detections was calculated for each object, taking into account the photometric redshift distribution 
(as our spectral coverage is incomplete), UV magnitude, and expected limiting flux for a likely MOSFIRE exposure. 
This exercise resulted in the final list of 16 candidates presented in Table 1.

\subsubsection{CLASH Lensing Sample}

Over the course of our November and March observations, we also targeted 3 candidates with a photometric redshift $z \geq 7.5$ from the CLASH HST survey
(GO 12065-12791 PI; Postman) as collated in \citet{Bradley2013a}.  Although these targets can only be surveyed individually, limiting 
our efficiency, as each is sampled with 8 HST filters at or longward of $\lambda = 7750$ \AA\, they have sharp redshift probability distributions, 
and well-determined UV slopes making them ideally suited for our new method.

We first targeted the $z \sim 7.9$ candidate in Abell 611 on November 5, 2013, securing 1.1 hours of on-source integration
in 0.8'' seeing. A further 1.5 hours of integration was possible on March 5, 2014.  In our March 2014 run,
we also targeted candidates in RXJ-1347 and MACS-0647 for 1.0 hours each.  Typical seeing conditions for the March nights 
were 0.60-0.65''. For all data, we used a 2.5'' AB dither pattern. Full details of the 3 targets are presented along
with our GOODS-S/UDF 2012 sample in Table 1. 

\subsubsection{Data reduction}

The data was reduced using the publicly available MOSFIRE data reduction pipeline\footnote{\rm https://code.google.com/p/mosfire/}.  
This pipeline first creates a median, cosmic-ray subtracted flat field image for each mask.  
Wavelength solutions for each slit are fit interactively for the central pixel in each slit, then propagated outwards to the 
slit edges to derive a full wavelength grid for each slit. The sky background is estimated as a function of wavelength 
and time using a series of B-splines and subtracted from each frame.  The nodded A-B frames are differenced, 
stacked, rectified and output for use along with inverse variance-weighted images used for error estimation.

Examination of our GOODS-S data revealed a gradual $0.6$'' ($\sim 3$ pixel) drift in 
the spatial direction over the course of our integration, which needed to be corrected for. 
The drift was tracked using a \jhst $\sim 19$ star 
conveniently placed on one slit.  The intensity of the star allowed us to follow the extinction for each frame and
eliminating those frames affected by significant extinction or drift, we secured 2.35 hours of useful exposure.
Due to this drift, the star on our original reduction displayed a somewhat greater FWHM 
of $\sim 1.2$'' than any of our individual exposures, which typically had a FWHM $\sim 0.8$''.
To correct this, we registered the relative positions of all frames by fitting a gaussian profile to star along the spatial
axis.  Given the drift over the entire exposure, we then arranged the frames into three separate 
groups, with the spatial positions in all frames consistent to within $\sim 1$ pixel.  Each of the three frame groups
was reduced individually using the same procedure described above.  To produce our final science stack, the three 
reductions were then shifted by the appropriate integer number of pixels and coadded, weighting by exposure time.  Using
this method, we were able to reduce the observed stellar FWHM to $\sim 1.0$''.
Our final mask reaches a median $5 \sigma$ limiting sensitivity between skylines of $\sim 7.0 \times 10^{-18}$ \fcgs. 
We note that approximately $\sim 33 \%$ of the Y-band spectral range is obscured by skylines at the MOSFIRE resolution of R $\sim 3380$
given our 0.7'' slit width.

\subsection{A New z=7.62 Lyman Alpha Emitting Galaxy}

We inspected the reduced, two-dimensional spectra of all 16 objects by eye to search for \lya\ emission.  From this, we 
were able to locate only a single candidate emission line.  Surprisingly, this emission line is located in one of our
faintest targets, UDF12-3313-6545 (first identified by \citealt{McLure2010a}, \citealt{Bouwens2011a}), with a measured
flux of $5.6 \times 10^{-18}$ \fcgs.  If the line is indeed \lya, the galaxy lies at a spectroscopic redshift of $z = 7.62$, 
making it a promising candidate for the most distant spectroscopically-confirmed galaxy to date.  
We present the relevant details and HST cutouts of this galaxy in Figure \ref{fig:udf_40242_sed}. 

Given the faintness (the emission line is detected at $4.0 \sigma$), line asymmetry, commonly
used to distinguish \lya, is not detectable.  However, the fact that the line displays a clear positive signal flanked by two negative peaks
indicates that the signal was present in both the A and B exposure positions.  Although the line has a surprisingly large rest-frame equivalent width
of $160 \pm 40$ \AA , this is comparable to some discovered in $z \sim 6$ \lya\ emitting galaxies \citep{Ouchi2010a}. Notably, the spectroscopic 
redshift lies well within to the $1 \sigma$ confidence interval of our derived photometric redshift distribution when line emission is accounted for, 
instilling further confidence in the redshift.

\subsection{Additional Data from Published Surveys}

In order to achieve the most up-to-date and precise measurement of the \lya\  opacity at 
$z \geq 6.5$, we have compiled a comprehensive sample of other near-infrared surveys for \lya\ at high redshift, which
we will utilize in our analysis. This includes our own prior work with Keck's NIRSPEC \citep{Schenker2012a}, as well as a number
of other surveys, using red-sensitive optical detectors on the VLT and Keck \citep{Fontana2010a,Pentericci2011a,Ono2012a,Pentericci2014a},
as well as independent work by Treu and collaborators using NIRSPEC \citep{Treu2012a} and MOSFIRE \citep{Treu2013a}.

In total, this literature sample comprises 83 $z \geq 6.5$ galaxies for which follow-up spectroscopy at various depths has
been attempted, plus an additional 19 from this work. To apply our method, we split this sample into 
two redshift bins centered at $z \sim 7$ and $z \sim 8$. 
The manner in which targets were assigned to each bin required careful consideration given the limited wavelength
response of each instrument with respect to the photometric redshift likelihood distribution $P(z)$. Rather than binning
on photometric redshift alone, we carefully considered the redshift range within which a null detection could be determined. If the
median redshift for which a null detection could be determined was less than (greater than) $z = 7.5$, we place the object
in the $z \sim 7 (8)$ bin.

\subsubsection{Monte Carlo simulation}

To predict the number of detections expected in an IGM with no additional opacity to \lya, we use a similar Monte Carlo
method to that developed in \citet{Schenker2012a}. This simulation has three key inputs for each object: flux limits 
as a function of wavelength from the spectroscopic observations, which also take into account the night sky emission, a photometric
redshift probability distribution, and a prediction for the IGM unprocessed \ewlya\ 
(and thus $f_{Ly\alpha}$) distribution. 

For objects observed in either this paper, or \citet{Schenker2012a}, flux limits  as a function of wavelength were calculated 
directly from the reduced spectra by computing the variance in an aperture of 10 \AA\ spectral extent.  For the data in 
\citet{Treu2012a} and \citet{Treu2013a} we rescaled our flux limits from NIRSPEC and MOSFIRE, respectively, to match 
the quoted limits in the paper for each object.  For \citet{Pentericci2011a,Ono2012a,Pentericci2014a}, we did the same 
with our LRIS limits, as presented in \citet{Schenker2012a}. 

We used the published photometry from each paper (and our own here) in conjunction 
with our photometric redshift code described previously to determine a photometric
redshift distribution for each object. The only exceptions are for the samples from \citet{Pentericci2011a} and \citet{Ono2012a},
for which either photometry or coordinates were not available. For these objects, we used the photometric
redshift distribution for ground-based $z$-drops from \citet{Ouchi2009a}. 

Finally, for the objects in our new MOSFIRE survey, we generated the prediction for the IGM unprocessed \ewlya\ distribution using 
the observed UV slope, as described in Section \ref{sec:analysis}.  Ideally, we would prefer to use this new method 
for all objects in the combined sample, in order to eliminate the potential bias of simply using
\muv\ as a predictor. However, with the exception of galaxies in the UDF 2012 field and CLASH lensed sample, the requisite 3 
infrared photometric data points longward of the Lyman break essential for achieving an accurate measure of $\beta$ are not available. 
Thus, for all other objects, we must predict \ewlya\ as a function of \muv\ from \citet{Treu2012a}, using the data presented in 
\citet{Stark2011a}. As an illustrative exercise,
we also generated a prediction for the IGM unprocessed \ewlya\ distribution of these objects using their \muv\ to calculate
a $\beta$ derived from the \muv -$\beta$ relation at $z \sim 7$ from \citet{Bouwens2013a}. Though not plotted, these results
are available in Table \ref{table:mc_results}.

With these inputs, we conduct a Monte Carlo simulation. In each trial for a given object, we draw a redshift from the 
photometric redshift distribution, and a \lya\ equivalent width from the \ewlya\ distribution. From
the observed photometry, this \ewlya\ is converted to a flux. We then sample the spectroscopic flux limit at the redshift drawn
to determine if the emission line would be observed at $\geq 5 \sigma$. This process is then repeated with $N = 10000$ trials
for each object.  

\subsubsection{Comparison between UV slope and UV luminosity predictions}

Before considering the total sample (i.e. including previous data from the literature), we compare the difference in
the expected \lya\ statistics for the high redshift sample using either \muv\ or $\beta$ as the basis for predicting the IGM unprocessed 
\ewlya\ distribution. Since only the UDF 2012 and CLASH surveys currently have accurate individual
measurements of $\beta$, this comparison can only be done for the 13 targets from our recent MOSFIRE survey.

The number of expected detections is compared in Figure \ref{fig:beta_muv_comp}. Although hampered by limited
statistics, the difference is still significant. Using  $\beta$ as a basis, we predict an average of 1.4 more detections than  
using \muv . This difference represents an important correction of a systematic error in the prior \xlya\ tests. Our new
results show that the \muv\ method, for this specific sample of 13 targets, significantly underpredicts the incidence of
IGM unprocessed \lya\ emission, which results in an overestimate of the IGM transmission.  The difference in 
predicted detections is dependent upon the properties of the sample considered
but, as the objects probed from the both the UDF and CLASH are intrinsically faint, with blue UV slopes, it is
not surprising that the difference is so great.  

This change in predicted \lya\ emission has important consequences for the transmission fraction of the IGM implying
a lower limit on the neutral fraction that is a factor 0.16 larger. Clearly, for a given survey, our new $\beta$ method 
for predicting the IGM unprocessed \ewlya\ distribution has significant advantages, and reduces a key systematic
error. This will be especially relevant for 
spectroscopic follow-up of the  HST Frontier Fields (GO:13496, PI: Lotz), and their parallels, given these fields 
will have full coverage with the same four WFC3/IR filters used for the UDF.

\subsubsection{Analysing the Entire Sample}

We can now combine the various subsamples, our own MOSFIRE survey and that from diverse sources in
the literature. The net result is a histogram of the number of $5 \sigma$ detections overall. These histograms
are displayed, both for the $z \sim 7$ and $z \sim 8$ samples in Figure \ref{fig:pred_det_all}; our new MOSFIRE
survey is, by design, more effective in constraining  the higher redshift limits.  We see evidence for
a moderate decline in the Lyman alpha fraction at $z \sim 7$ and a continued sharper decline at higher redshift. Note
that our new spectroscopic confirmation at $z = 7.62$ is not included as a detection as it lies below the a priori 
$5 \sigma$ flux limit.  

As an illustration, we can convert these observational-based results to an IGM extinction of \lya\ by adopting the
model used in \citet{Schenker2012a} appropriate for patchy reionization.  In this model, the IGM is partially opaque, 
such that \lya\ escapes to the observer unattenuated from a fraction, $f$, of galaxies, while it is completely extinguished 
by the IGM in a fraction $1-f$. Given the histogram of expected detections and the number actually observed, we infer a 
probability distribution for this transmission fraction, $f$. At  $z \sim 7$, we find $f = 0.6 \pm 0.15$, and at $z \sim 8$, 
a $1\sigma$ upper limit of $f < 0.19$. The full results  can be found in Table \ref{table:mc_results} and are plotted
along with the lower redshift data on \xlya\ in Figure \ref{fig:xlya_vs_z}.

Discussing the uncertainties in the transformation from transmission fraction to \xhi\ is beyond the scope of the present
paper. However, clearly this conversion is dependent upon a number of physical parameters, some internal to the 
galaxy, and others from the IGM state itself. These include the velocity offset of \lya\ from the galaxy's systemic velocity 
(e.g. \citealt{Hashimoto2013a,Schenker2013b}), the ionizing photon escape fraction \citep{Dijkstra2014a}, as well as the 
possible presence of optically thick absorption systems \citep{Bolton2013a}. Until the theoretical models converge, 
and/or observations of these key quantities are available, absolute measures of the neutral fraction will still be subject to systematic
errors. Nonetheless, we have demonstrated substantial observational progress with our new
survey and improved methodology, reducing one of the key systematic errors. Using the models of \citet{McQuinn2007a} here to 
provide an estimate of \xhi, we find
\xhi\ $=  0.34^{+0.09}_{-0.12}$ at $z \sim 7$, and \xhi\ $> 0.65$ at $z \sim 8$.

\section{Conclusions}

Using our sample of 451 $3 < z < 6$ spectroscopically followed-up Lyman break galaxies, we demonstrate an improved
correlation between the ultraviolet continuum slope of a galaxy, $\beta$, and its \lya\ emission strength.  Given the
availability of deep WFC3 photometry for both the GOODS-N and S fields, this progress follows measurements
for many individual galaxies in this redshift range, rather than via stacked or averaged UV slopes, as in earlier
work \citep{Shapley2003a,Stark2010a}. 

We demonstrate that this correlation with the presence of \lya\ is stronger and more physically-motivated than that based on the UV
luminosity and thus provides a natural basis for an improved model for the \lya\ fraction test, now widely used to measure
the evolving neutrality of the $z > 6.5$ IGM. We demonstrate the benefits of this new model using a new MOSFIRE
spectroscopic survey of $7<z<8$ targets from the Ultra Deep Field 2012 catalog and CLASH lensing survey, and 
combine this with data at these redshifts already published
in the literature.  As a result we present the implications of the most comprehensive search for \lya\ emission 
at $z \simeq 8$ to date, confirming once again important evidence that cosmic reionization ended at redshifts $z \simeq 6.5$.

As a byproduct we also present the $4.0\sigma$ confirmation of \lya\ in a galaxy at $z = 7.62$, likely the 
most distant spectroscopically-confirmed galaxy.

\acknowledgements

We thank Chuck Steidel and Ian McLain for their sterling efforts in developing the highly successful MOSFIRE 
instrument.  We also wish to recognize and acknowledge the very signiﬁcant cultural role and 
reverence that the summit of Mauna Kea has always had within the indigenous Hawaiian 
community. We are most fortunate to have the opportunity to conduct observations from 
this mountain.

\begin{deluxetable*}{lcccccc}
\tablecolumns{7}
\tablewidth{0pt}
\tablecaption{\bf Summary of MOSFIRE survey for Ly$\alpha$}
\tablehead{\colhead{Name} & \colhead{RA} & \colhead{Dec} & \colhead{\jhst} &\colhead{z$_{phot}$} & \colhead{t$_{exp}$ [hr]} & \colhead{$5\sigma$ EW limit}}
\medskip
\startdata

A611-0193       & 8:01:00.32    & 36:04:24.3    & 26.0  & 7.9 & 2.6 & 12.9\\
MACS0647-1411   & 6:47:40.91    & 70:14:41.6    & 26.1  & 7.6 & 1.0 & 20.7\\
RXJ1347-0943    & 13:47:33.90   & -11:45:09.4   & 26.4  & 7.5 & 1.0 & 27.7\\
\hline
\\
CANDY-2272447364        & 3:32:27.24    & -27:47:36.4   & 27.1  & 9.1 & 2.35 & 60\\
CANDY-2243349150        & 3:32:24.33    & -27:49:15.0   & 27.2  & 8.5 & 2.35 & 57\\
GS-2098-8535    & 3:32:20.98    & -27:48:53.5   & 27.2  & 9.7 & 2.35 & 64\\
GS-2533-8541    & 3:32:25.33    & -27:48:54.1   & 26.4  & 7.6 & 2.35 & 30\\
GS-2779-5141    & 3:32:27.79    & -27:45:14.1   & 27.2  & 9.6 & 2.35 & 64\\
GS-4402-7273    & 3:32:44.02    & -27:47:27.3   & 26.9  & 7.4 & 2.35 & 50\\
UDF12-3313-6545 & 3:32:33.13    & -27:46:54.5   & 28.5  & 7.4 & 2.35 & 200\\
UDF12-3722-8061 & 3:32:37.22    & -27:48:06.1   & 27.8  & 7.7 & 2.35 & 110\\
UDF12-3846-7326 & 3:32:38.46    & -27:47:32.6   & 29.0 & 7.0 & 2.35 & 330\\
UDF12-3880-7072 & 3:32:38.80    & -27:47:07.2   & 26.9  & 7.5 & 2.35 & 220\\
UDF12-3939-7040 & 3:32:39.39    & -27:47:04.0   & 28.6  & 7.7 & 2.35 & 180\\
UDF12-3762-6011 & 3:32:37.62    & -27:46:01.1   & 28.4  & 8.1 & 2.35 & 150\\
UDF12-3813-5540 & 3:32:38.13    & -27:45:54.0   & 28.2  & 8.2 & 2.35 & 70\\
UDF12-4256-7314 & 3:32:42.56    & -27:47:31.4   & 27.3  & 7.1 & 2.35 & 50\\
UDF12-4308-6277 & 3:32:43.08    & -27:46:27.7   & 28.5  & 8.0 & 2.35 & 200\\
UDF12-4470-6443 & 3:32:44.70    & -27:46:44.3   & 27.3  & 7.7 & 2.35 & 70

 \smallskip
\enddata
\tablecomments{\label{table:xlya_targ} MOSFIRE survey targets. 5$\sigma$ limiting sensitivities are
calculated using the median limiting flux limit between skylines, and assuming a spectroscopic redshift 
of $z = 7.7$.}

\end{deluxetable*}

\begin{deluxetable*}{lccccc}
\tablecolumns{6}
\tablewidth{0pt}
\tablecaption{\bf Summary of MOSFIRE survey for Ly$\alpha$}
\tablehead{\colhead{Survey} & \colhead{Observed} & \colhead{$5\sigma$ Detections} & \colhead{Transmission fraction ($f$)} &\colhead{\xhi}}
\medskip
\startdata

This work MOSFIRE \muv & 19 & 0 & $ < 0.47 $ & $ > 0.43 $ \\
This work MOSFIRE $\beta$ & 19 & 0 & $ < 0.36 $ & $ > 0.50 $ \\
Composite $z \sim 7$ & 72 & 11 & $ 0.51_{-0.11}^{+0.14} $ & $ 0.40_{-0.10}^{+0.08} $ \\
Composite $z \sim 8$ & 27 & 0 & $ < 0.19 $ & $ > 0.65 $ \\
\hline
\\
Composite $z \sim 7$\tablenotemark{a} & 72 & 11 & $ 0.34_{-0.06}^{+0.13} $ & $ 0.51_{-0.09}^{+0.05} $ \\
Composite $z \sim 8$\tablenotemark{a} & 27 & 0 & $ < 0.15 $ & $ > 0.68 $ 

\smallskip
\enddata
\tablenotetext{a}{For these results, where individual UV slopes are not available, we instead use individual
values of \muv\ to predict a value of $\beta$, which in turn is used to generate the IGM unprocessed \ewlya\ distribution.}
\tablecomments{\label{table:mc_results} Monte Carlo results for transmission fraction, $f$, and \xhi. }

\end{deluxetable*}

\begin{figure*}
\begin{center}

\includegraphics[width=0.6\textwidth]{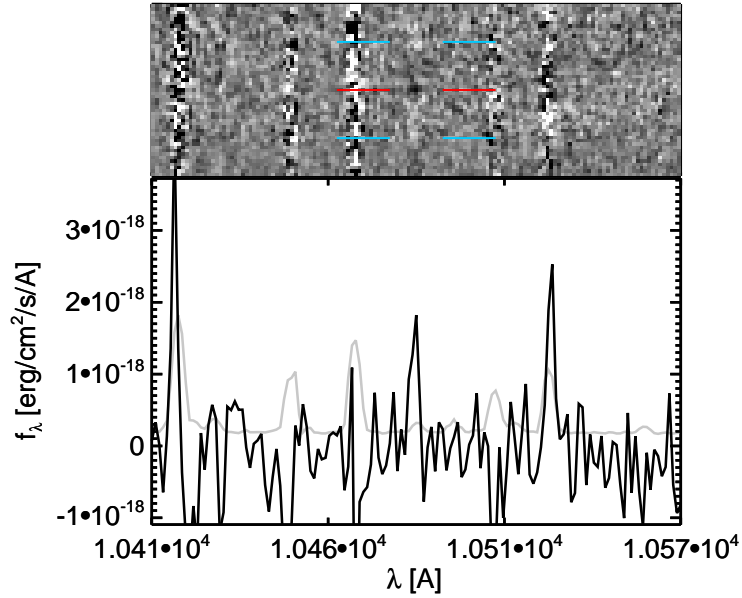}

\caption{\label{fig:spec_2d} MOSFIRE observations of our lone target with visible line emission, UDF12 40242. The full
two dimensional spectrum is shown at top, with the one dimensional spectrum, plotted at the bottom, along with the 
error spectrum in grey.  Given our A$-$B reduction strategy (described in Section 2), our 2D spectrum shows the expected
positive signal (red line) flanked by two negative signals (blue lines), each separated by the amplitude of the dither pattern.}
\end{center}
\end{figure*}

\begin{figure*}
\begin{center}

\includegraphics[width=0.9\textwidth]{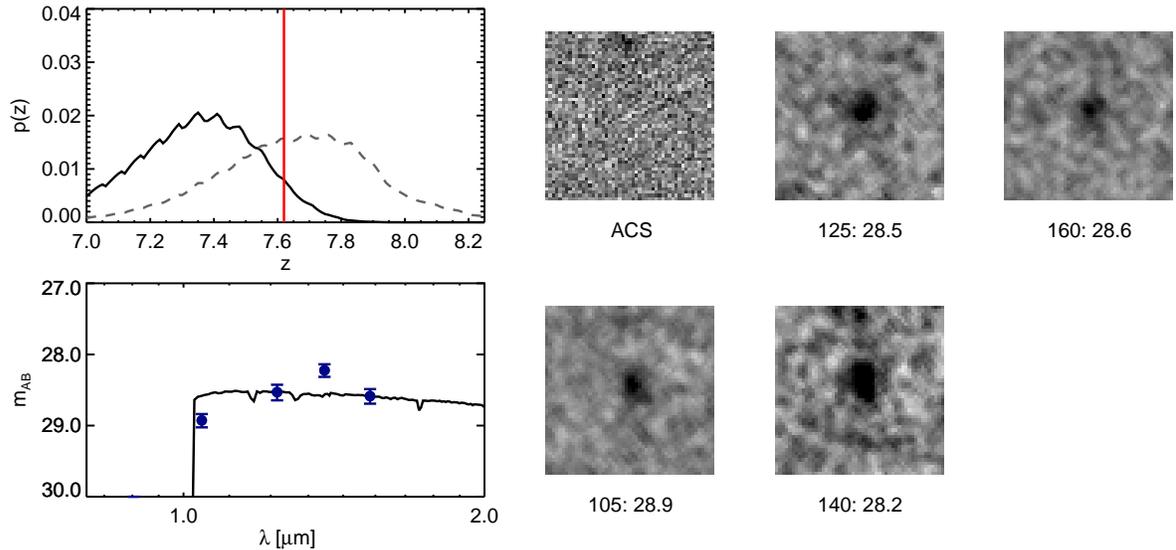}

\caption{\label{fig:udf_40242_sed} Right: 1.5 arcsecond diameter cutouts and total magnitudes (see data section) of our MOSFIRE 
target, UDF12 40242.  As expected for a $z > 7$ candidate, the object is not formally detected in a stack of the optical data. Left, Top: 
Photometric redshift probability distribution function (pdf) for our target, with the actual spectroscopic redshift, $z=7.62$ denoted in red.  The solid black curve displays the pdf from the raw photometry, while the dashed grey curve shows the pdf after the observed
MOSFIRE line flux has been subtracted from the \yhst\ data point. Left, Bottom: Best fitting SED for the galaxy, along with HST
WFC3 Photometry.}
\end{center}
\end{figure*}

\begin{figure*}
\begin{center}

\includegraphics[width=0.3\textwidth]{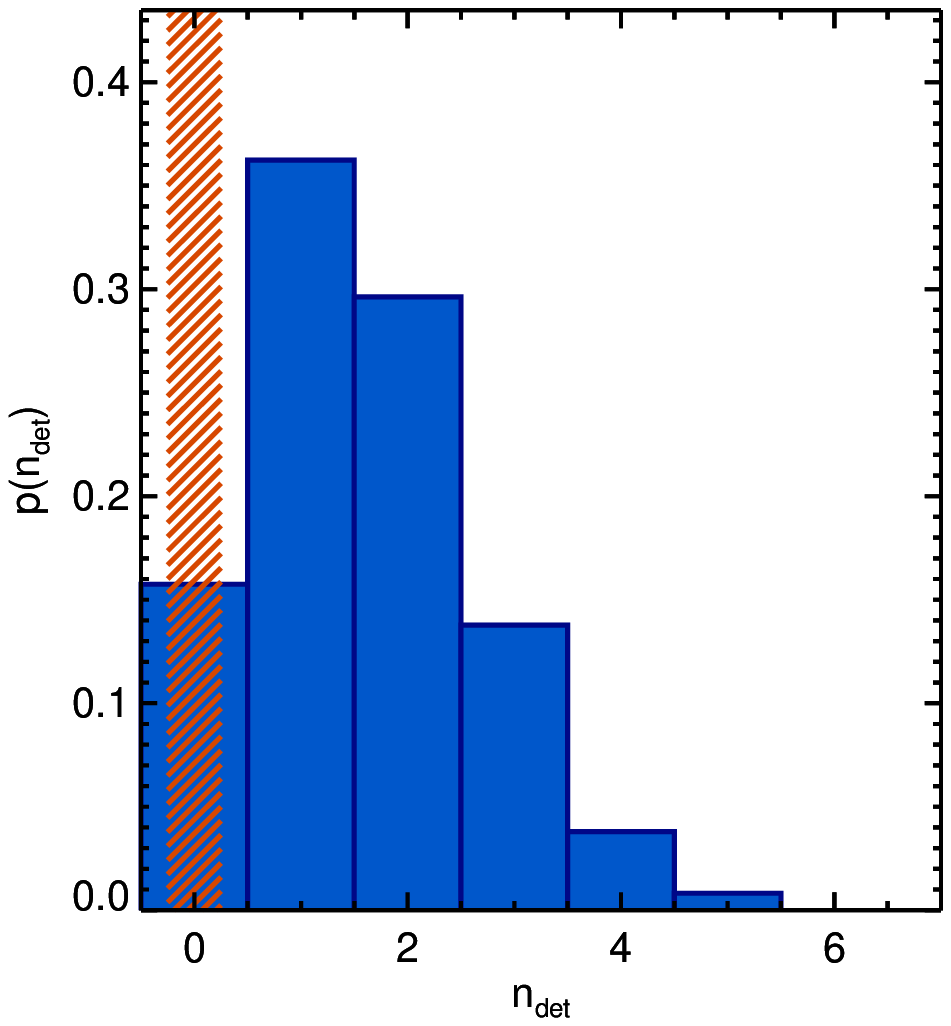}
\includegraphics[width=0.3\textwidth]{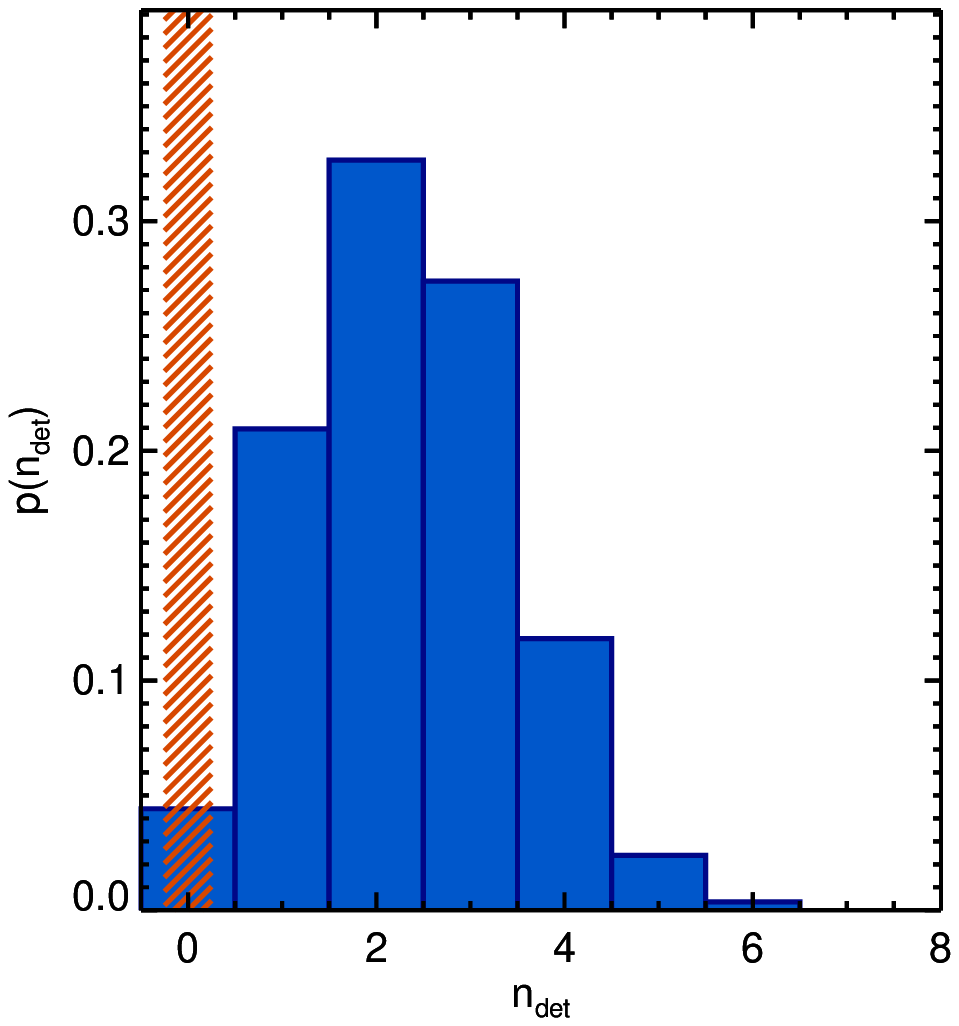}

\caption{\label{fig:beta_muv_comp} Left: Predicted number of detections for our MOSFIRE survey using the 
\ewlya\ probability distribution from \citet{Treu2012a}, which uses \muv\ as the predictor. The observed number
of 0 detections is indicated by the red crosshatch. Right: Predicted number of detections for same survey, but using 
$\beta$ as the predictor for \ewlya,
as outlined in Section \ref{sec:analysis}. In this case, the average number of expected detections is increased by a 
factor of 0.4, highlighting the importance of using a model that accurately predicts the IGM unprocessed equivalent width
distribution. The equivalent upper limit on the transmission fraction is also decreased by a factor of 0.23.}

\end{center}
\end{figure*}

\begin{figure*}

\begin{center}
\includegraphics[width=0.65\textwidth]{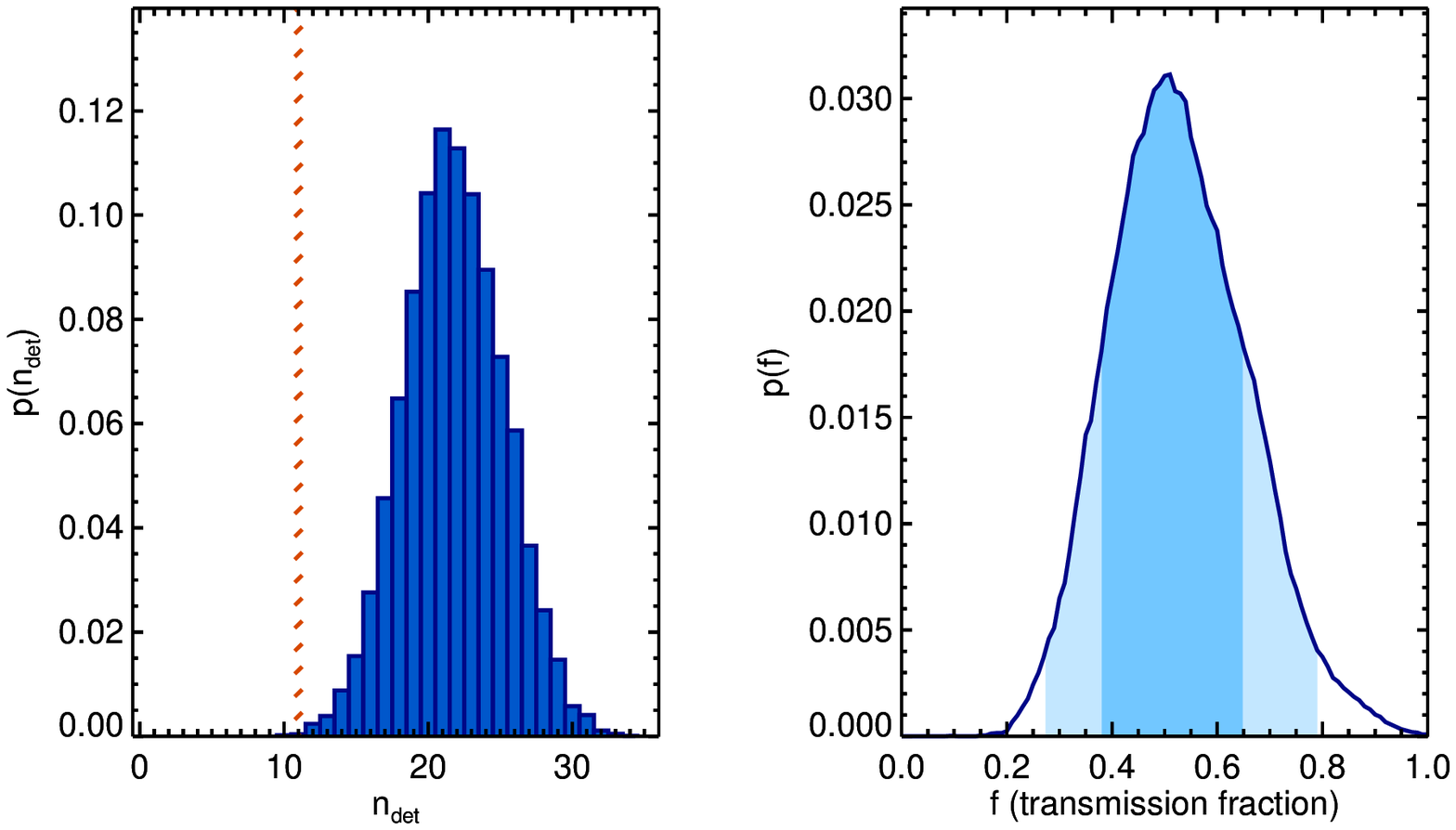}
\includegraphics[width=0.65\textwidth]{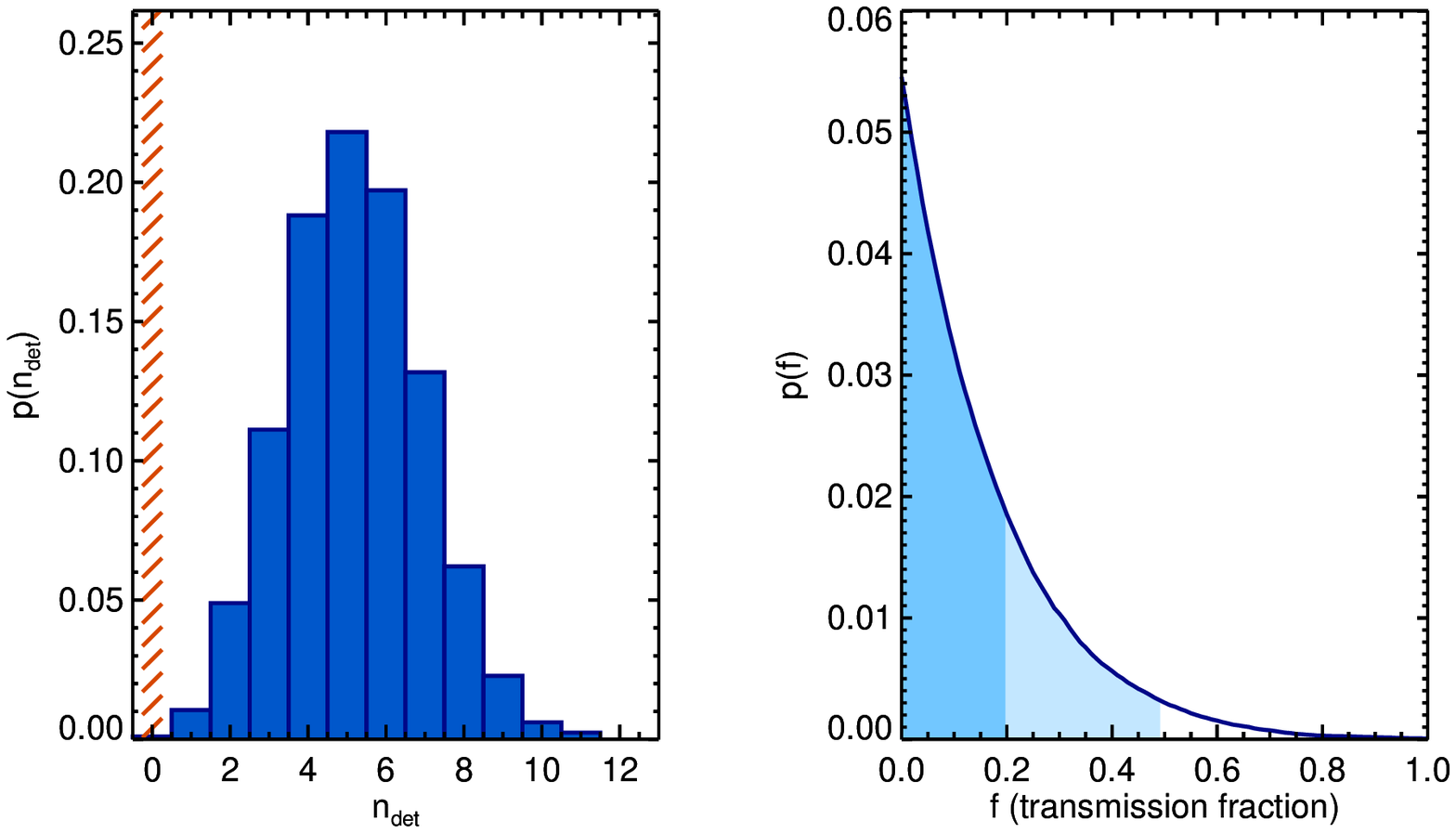}

\caption{\label{fig:pred_det_all} Results from our new MOSFIRE campaign, combined with data from the literature. 
Top left: Histogram of expected 5$\sigma$ detections of
\lya, computed using the Monte Carlo method described in the text for our $z \sim 7$ sample. The red crosshatches denote the 
combined number of detections observed in all surveys. Top right: Given our predicted and observed number of detections, the constraints on the average extinction fraction of \lya, assuming a patchy opacity. Dark blue and light blue shading
encompass the 1$\sigma$ and 2$\sigma$ confidence intervals, respectively. Bottom left and right:
Same as above, but for our $z \sim 8$ sample.}
\end{center}
\end{figure*}

\begin{figure*}
\begin{center}
\includegraphics[width=0.5\textwidth]{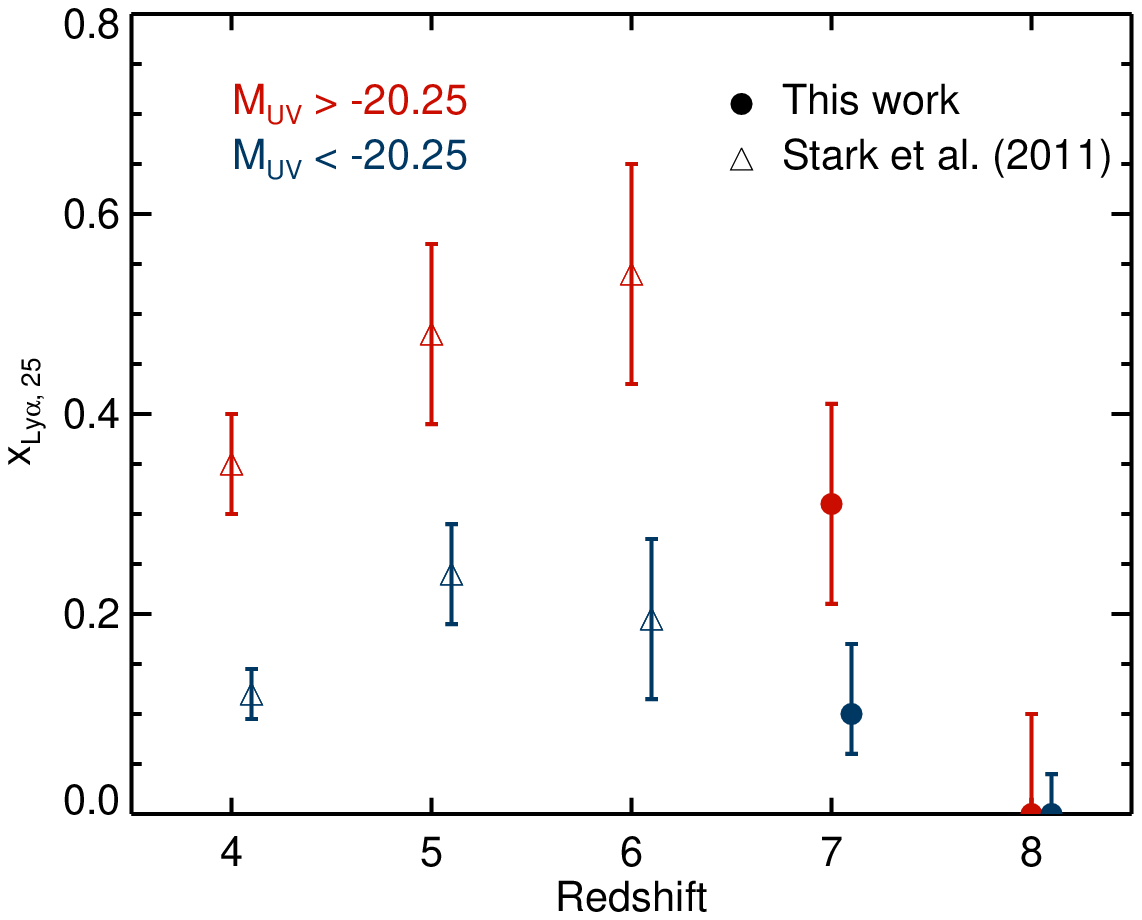}

\caption{\label{fig:xlya_vs_z} The fraction of Lyman break galaxies that display \lya\ in emission at
an EW $\geq 25$ \AA, plotted as a function of redshift. The values at $z = 7$ and 8 reflect differential measurements
with the data at $z = 6$, as described in the text. Thus, these data points and errors are simply the convolution of the 
\xlya\ PDF at $z = 6$ and the transmission fraction PDF at $z = 7$ and 8.}
\end{center}
\end{figure*}

\appendix

\section{A. Models for p(EW|$\beta$)}

The maximum likelihoods inferred from each of the four distributions are noted in Table \ref{table:distributions}.  These
results demonstrate that the lognormal distribution provides the best fit to the available data - its likelihood surpasses
that of any other model by two orders of magnitude. Thus, we use this distribution as the basis for the more general
form of $p(EW|\beta)$ we consider next.

\section{B. Results of full modeling procedure}

For reference, and such that they are available for use in future work, we list here the final values for our 
generalized lognormal fit to the \ewlya\ distribution at $3 < z < 6$. They are $\mu_a = 3.0^{+0.125}_{-0.25}$,
$\mu_s = -1.125 \pm 0.25$, $\sigma = 1.3 \pm 0.1$, and $A_{em} = 1.0^{+0.0}_{-0.05}$. We also provide
a plot of the posterior probability distribution below so the reader is able to appreciate the
sometimes non-negligible covariances between parameters.

\begin{figure*}
\begin{center}

\includegraphics[width=0.8\textwidth]{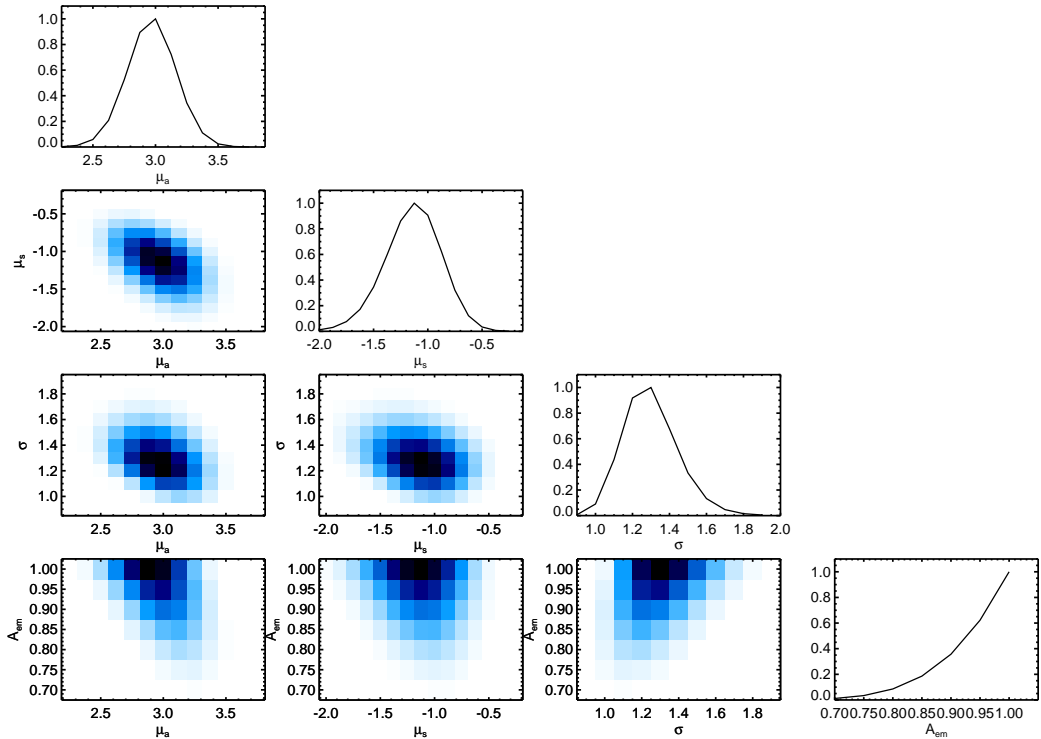}

\caption{\label{fig:all_beta_like}  Posterior probability distribution for our full model, $p$(\ewlya|$\beta$). Shaded
plots represent the posterior PDF marginalized over all but the two variables labeling the axes, while line plots
are marginalized over all but one variable. Thus, the one dimensional PDFs for each variable, from which
we quote our error bars, can be read off along the diagonal. }
\end{center}
\end{figure*}

\begin{deluxetable*}{lcccc}
\tablecolumns{5}
\tablewidth{0pt}
\tablecaption{\bf Ly$\alpha$ functional forms}
\tablehead{\colhead{Distribution name} & \colhead{Equation} &\colhead{Free parameters} & \colhead{Reference} 
& \colhead{Log$_{10}$ max likelihood}}
\medskip
\startdata

Lognormal  &  $ A_{em} \frac{1}{\sqrt{2\pi}\sigma \mathrm{EW}} exp(-(\mathrm{ln(EW)-\mu})^2 / 2 \sigma^2)  $ & $A_{em}$, $\mu$, $\sigma$  & This work & -75.4 \\
Half-gaussian  & $ \frac{1}{\sqrt{2\pi}\sigma} exp(-(\mathrm{EW}-\mu)^2 / 2 \sigma^2)  $  & $A_{em}$, $\sigma$ & \cite{Treu2012a} & -80.7\\
' w/ high-EW tail & $ A_{em,g}  \frac{1}{\sqrt{2\pi}\sigma} exp(-(\mathrm{EW}-\mu)^2 / 2 \sigma^2) + A_{em,c}$  & $A_{em,g}$, $A_{em,c}$, $\mu$, $\sigma$ & \cite{Pentericci2011a} & - 80.8\\
Declining exponential & $A_{em}  exp(-EW / EW_{0}) $ & $A_{em}$, $EW_{0}$ & \cite{Dijkstra2011a} & -77.9 \\
 \smallskip
\enddata
\tablecomments{\label{table:distributions} List of the mathematical distributions used to fit the \ewlya distribution at $\beta \sim -2.3$,
and the calculated maximum likelihoods. Having a maximum likelihood two orders of magnitude greater than any other distribution
considered demonstrates the lognormal distribution provides the best fit.}
\tablenotetext{a}{This is a note}

\end{deluxetable*}

\end{document}